\documentclass[conference]{IEEEtran}

\ifCLASSINFOpdf
  \usepackage[pdftex]{graphicx}
\else
  \usepackage[dvips]{graphicx}
\fi

\usepackage{setspace}
\usepackage[T1]{fontenc}
\usepackage[utf8]{inputenc}
\usepackage{tabularx}
\usepackage{color}
\usepackage{url}
\usepackage{units}
\usepackage{xspace}
\usepackage[table,hyperref,x11names]{xcolor}
\usepackage{balance}
\usepackage{array,multirow}

\usepackage{siunitx}
\usepackage{fixltx2e}
\usepackage{makeidx}

\usepackage{version}
\includeversion{mynote}
\usepackage{comment}

\usepackage[table,hyperref,x11names]{xcolor}
\usepackage[colorlinks=true,urlcolor=SteelBlue4,linkcolor=Firebrick4]{hyperref}

\usepackage[lined,ruled]{algorithm2e}
\usepackage[noend]{algpseudocode}
\usepackage{varwidth}
\usepackage{amsmath}

\usepackage[hide]{chato-notes}

\newtheorem{theo}{Theorem} 

\newtheorem{lem}{Lemma}
 
\newenvironment{proofm}{\noindent\emph{Proof:}}{\hfill$\Box$} 
 
\algrenewcommand\alglinenumber[1]{\footnotesize #1:}
\makeatletter
\algrenewcommand\ALG@beginalgorithmic{\footnotesize}
\makeatother

\algrenewcommand\textproc{}
\SetAlFnt{\small}
\SetAlCapFnt{\large}
\SetAlCapNameFnt{\large}
\SetFuncSty{texttt} 
\SetArgSty{texttt}
\SetProcNameSty{texttt}
\SetProcArgSty{texttt}
\SetDataSty{texttt}
\SetAlFnt{\scriptsize}
\SetAlCapFnt{\scriptsize}
\SetAlCapNameFnt{\scriptsize}

\newcommand{\coloredtext}[1]{#1\xspace}

\newcommand{\ofpacketin}{\texttt{PACKET\_IN}\xspace}
\newcommand{\offlowmod}{\texttt{FLOW\_MOD}\xspace}

\newcommand{\fourxspace}{\xspace\xspace\xspace\xspace}

\usepackage{amssymb}
\usepackage{pifont}

\usepackage{pgffor}

\begin{document}
\title{
The KISS principle in Software-Defined Networking:\\An architecture
for Keeping It Simple and Secure
}
\thanks{This work is partially supported by the University of Luxembourg - SnT and by the Fonds National de la Recherche Luxembourg (FNR) through PEARL grant FNR/P14/8149128.}

\author{\IEEEauthorblockN{Diego Kreutz, Jiangshan Yu, Paulo Esteves-Verissimo,}
\IEEEauthorblockA{SnT, University of Luxembourg, Luxembourg\\
\{Diego.Kreutz,Jiangshan.Yu,Paulo.Verissimo\}@uni.lu}

\and
\IEEEauthorblockN{C\'{a}tia Magalh\~{a}es, Fernando M. V. Ramos}
\IEEEauthorblockA{LaSIGE/FCUL, University of Lisboa, Portugal\\
catiamagalhaes27@gmail.com, fvramos@ciencias.ulisboa.pt}
}

\maketitle
\thispagestyle{plain} \pagestyle{plain}

\begin{abstract}

Security is an increasingly fundamental requirement in
Software-Defined Networking (SDN).
However, the pace of adoption of secure mechanisms has been slow,
which we estimate to be a consequence of the performance overhead of
traditional solutions and of the complexity of the support
infrastructure required.
As a first step to addressing these problems, we propose a modular
secure SDN control plane communications architecture, KISS, with
innovative solutions in the context of key distribution and secure
channel support.
A comparative analysis of the performance impact of essential security
primitives guided our selection of basic primitives for KISS.  We
further propose iDVV, the integrated device verification value, a
deterministic but indistinguishable-from-random secret code generation
protocol, allowing the local but synchronized generation/verification
of keys at both ends of the channel, even on a per-message basis.
iDVV is expected to give an important contribution both to the
robustness and simplification of the authentication and secure
communication problems in SDN.

We show that our solution, while offering the same security properties, outperforms reference alternatives, with performance improvements up to 30\% over OpenSSL, and improvement in robustness based on a code footprint one order of magnitude smaller.
Finally, we also prove and test randomness of the proposed algorithms.

\end{abstract}


\IEEEpeerreviewmaketitle

\section{Introduction}

In Software-Defined Networking (SDN), network control is separated
from the forwarding devices and logically centralised in a controller.
This separation is achieved by means of a protocol (typically,
OpenFlow) that enables the SDN controller to remotely populate the
forwarding tables of network switches.  The OpenFlow standard includes
Transport Layer Security (TLS)~\cite{rfc5246} as an \emph{optional} security feature for
authenticating forwarding devices and controllers and for encrypting
the communication channel.  However, to date most reported deployments
still use TCP for control traffic, and SDN controllers and switching
hardware with TLS support are still rare~\cite{kreutz2014sdnsurvey}.
Moreover, most deployments communicate control plane traffic in-band
with data traffic to reduce the infrastructure required, making
control plane communication vulnerable to different
attacks~\cite{kreutz2014sdnsurvey}.  For instance, a single malicious
forwarding device can intercept control traffic and become a dangerous
threat to the SDN infrastructure~\cite{antikainen2014spook}.

Four fundamental issues can slow down the rate of adoption of secure
mechanisms in SDN.  First, securing communications has a
non-negligible cost in terms of increased communications latency and
reduced performance.  Several recent studies have analysed this
overhead in various
contexts~\cite{naylor2014https,6129420,5702353}.
Second, the computing capabilities of commodity switches are typically
weak.  The typical SDN switch (e.g.~\cite{hp2013h3s,Advantech2015fin,nexcom2015nsa})
is equipped with a single or dual-core CPU running at approximately
1GHz, which compares unfavourably with the multi-core CPUs found in
typical commodity servers.  Imposing the additional cost of TLS to
these computing-constrained networking devices is a problem.  
Third, poor choice of cryptographic primitive implementations can also 
have a significant impact on the performance of the control plane
communications handled by the controller.  Finally, the Public Key
Infrastructure (PKI) on which TLS relies is complex and thus
vulnerability prone~\cite{Wazan2013complexPKI,PKI_Chapter17}, opening
a large surface for successful
attacks~\cite{Meulen2013DigiNotar}. 

In order to meet these challenges, we propose a modular secure SDN
control plane communications architecture KISS (Section~\ref{sec:arch}), 
which aims to increase the robustness of control communications
whilst enhancing their performance, by decreasing the complexity of 
the support infrastructure, as an alternative to current approaches 
based on classic configura- tions of TLS and PKI.


A core novel component of our architecture is 
the integrated device verification value (iDVV), a deterministic but
indistinguishable-from-random secret code generation protocol 
(Section~\ref{sec:simplicity}).  The
concept was inspired by the iCVVs (integrated card verification
values) used in credit cards to authenticate and authorize
transactions in a secure and inexpensive way.  We develop and extend
the idea for SDN, proposing a flexible method of generating iDVVs by
adapting proven one-time password-like techniques. iDVV codes allow the
safe decentralized generation/verification of keys at both ends of the
channel, at will, even on a per-message basis.

To understand and minimize the cost of security, 
we quantify (Section~\ref{performance}) the impact of secure
primitives on the performance and scalability of control plane
communications, through a compared study of different implementations
of TCP vs. TLS, complemented by a deeper study of underlying hashing
and message authentication code (MAC) primitives.  Those experiments
confirm our intuition that the choice of protocols and primitives used
in secure communication may well be one strong reason behind the slow
adoption of these mechanisms in SDN.
This in-depth study lead to the selection of the
NaCl cryptographic library~\cite{Bernstein2012TSI}, and the best performing MAC and hash
primitives --- Poly1305 and SHA512 OpenSSL -- as the baseline secure
channel technologies for KISS.

iDVVs team-up with NaCl, in order to safely replace the cryptographic
primitives and key-exchange protocols and key derivation functions
commonly used in TLS.  As a result, the NaCl-iDVV compound, while
achieving the same functional level of security, is simpler,
potentially leading to a higher level of implementation robustness by
vulnerability reduction. In fact, we estimate the proposed security
architecture footprint to be smaller than TLS-PKI alternatives with
traditional protocols, by an order of magnitude, in terms of the
number of lines of code (LOC). Such a differential also points to
reducing the cyclomatic complexity.  These metrics are typically used
to assess the robustness and estimate verifiability of software systems.

Furthermore, in Section~\ref{sec:eval} we evaluate the iDVV
design in terms of performance, security and randomness. Key
generation latency of iDVV compares very favourably with common
implementations of key derivation functions. On the security side, we
prove the indistinguishability-from-random and determinism of the iDVV
generator. Finally, the iDVV successfully passed several empirical
randomness tests, further confirming its
indistinguishability-from-random, and showing its suitability for
highly-robust key generation.
We end the paper with a discussion and some pointers to further work.

\section{KISS architecture}
\label{sec:arch}

\begin{figure}[ht]
   \centering
   \includegraphics[width=0.95\columnwidth]{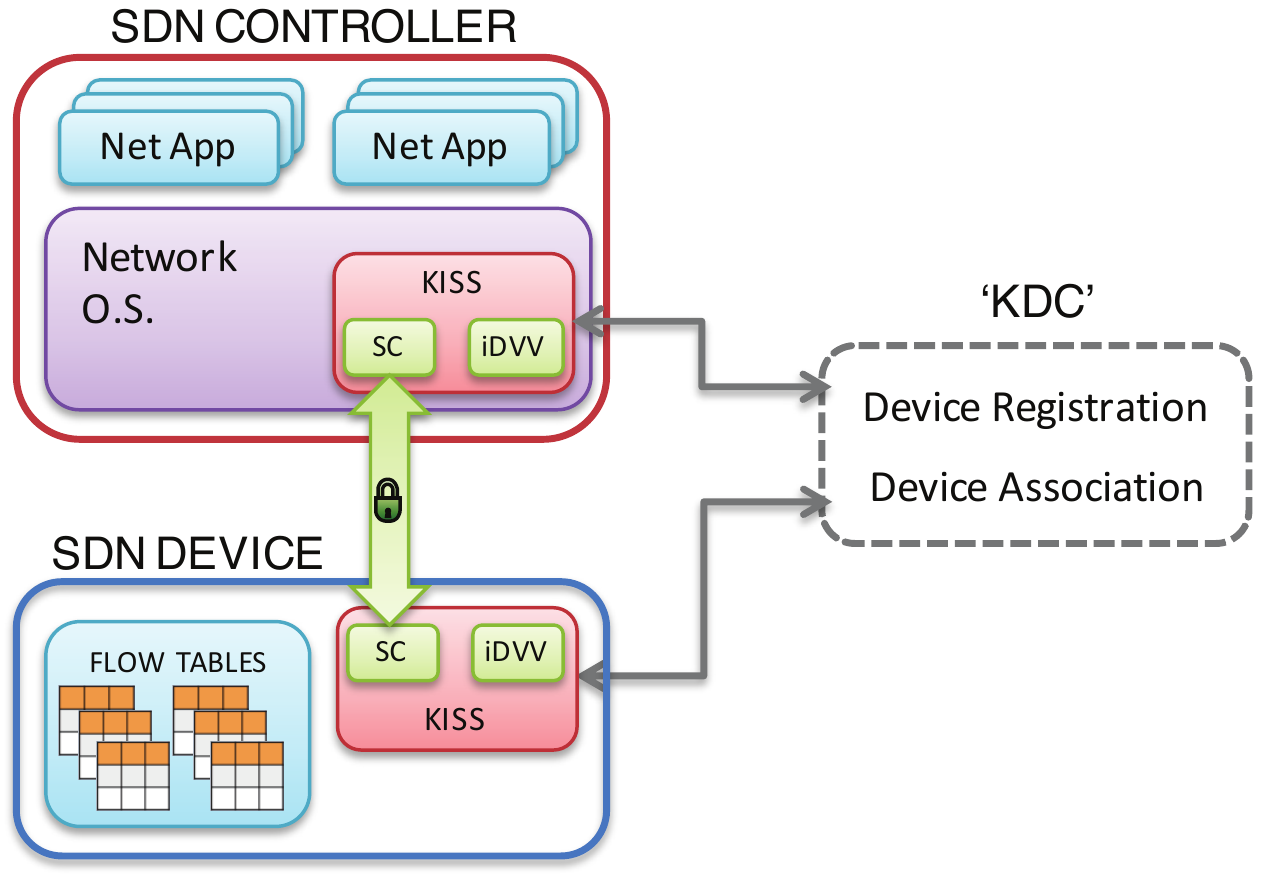}
   \caption{\coloredtext{General architecture}}
   \label{fig:architecture}
\end{figure}

In this section we present our proposal of KISS, a modular secure control
plane communications architecture for SDN offering
alternatives to classic configurations of secure channel and
authentication protocols and subsystems followed in TLS and PKI.
We assume a typical SDN architecture, as illustrated in
Figure~\ref{fig:architecture}, composed of controllers and forwarding
devices. We further assume that device registration and association
services are in place. For lack of space, we do not discuss them in
detail, but for self-containment, we discuss some properties and their
interface below.

The two components encapsulated by the KISS boxes are the crucial
components of the architecture, and the main subject of our study: a
secure channel protocol suite, composed of a judicious choice of
state-of-the-art mechanisms and protocols, which we dub SC for
convenience of description, and a novel deterministic but
indistinguishable-from-random secret code generation protocol, which
we call iDVV (integrated device verification value).

We have considered using TLS implementations (e.g. OpenSSL) as the baseline
protocol for SC.  However, the experiments in Section~\ref{performance} have alerted
us to: the sheer performance cost of cryptographic communication; and
the further impact of sub-optimal choices of cryptographic primitives.
This motivated us to adopt NaCl~\cite{Bernstein2012TSI},  
a high performance yet secure cryptographic library, as the substrate of SC, 
complemented by the MAC and strong hash primitives with best performance 
according to our experiments -- Poly1305 and SHA512 OpenSSL. SHA-512 is 
used by the iDVV generator while Poly1305 is a fast MAC algorithm.

The iDVV, a novel component we propose, helps to further
enhance the security of SC, through strong crypto material
generated at a low cost (e.g. one-time keys, per-message
authentication and authorization codes) to be used by NaCl
ciphers. The indistinguishability-from-random allied to the
determinism allow the safe decentralized generation/verification of
per-message keys at both ends of the channel.

\subsection{System and threat model}
\label{sec:sysmod}

For simplicity and without loss of generality, we assume that the
controllers and forwarding devices are registered and associated
through a secure and robust key distribution service provided by a key
distribution center (KDC), which for space reasons is out of the scope
of this paper, but can be readily secured by state-of-the-art KDCs
like Kerberos Key Distribution Center~\cite{neuman1994kerberos}.


The device registration process is by default invoked by network
administrators to the KDC, to register new devices.  In result of
device registration, the device and the KDC securely share a symmetric
key. We denote $K_{kc}$ the shared key between the KDC authority and a
registered controller, and $K_{kf}$ the shared key between the KDC
authority and a registered forwarding device.

Registered controllers and forwarding devices must be securely
associated, also through the KDC authority, as a pre-condition to
communicate securely.  The most common case is a forwarding device
$f_i$ requesting an association to a controller $c_j$, through the
KDC.  After associating, a controller and a forwarding device share
two symmetric secrets (of size 256 bits), namely a $seed_{ij}$ and a
$key_{ij}$. The key is generated by the KDC and the seed is generated
by the KDC in cooperation with the controller.  These secrets will be used to
bootstrap the iDVV module, as we discuss ahead.


As threat model, we consider a Dolev-Yao style attacker, who has a
complete control of the network, namely the attacker logs all
messages, and can arbitrarily delay, drop, re-order, insert, or modify
messages.
In addition, this strong attacker is able to compromise any network 
device (e.g. a controller or a forwarding device) at any time.
We assume the security of the used cryptographic primitives, including
MAC (i.e. Poly1305), hash function (i.e. SHA-512), and symmetric
encryption algorithm (e.g. AES). We will prove the security of the
iDVV codes in Section~\ref{sec:eval}. We also assume that the device
registration and association services can rely on robust pseudo-random
number generators.

\subsection{Security goals}
\label{sec:secgoals}

The main goal of KISS is to provide security properties including
authenticity, integrity, and confidentiality for control plane
communications, while minimizing cost and complexity.

The secure communication between participants can be easily guaranteed
when a secure encryption algorithm is used, as long as the shared
secret key is kept secure. 
To provide a robust SDN system, we focus on advanced security guarantees for the situation when the shared key is exposed to an attacker, as this might happen in practice. In particular, if an attacker has compromised a device and learnt its shared keys, then we are aiming at providing ``perfect forward secrecy'' (PFS) of communications. That is, the secrecy of a device's past communications should be protected when the device is compromised and its shared keys are exposed to an attacker. It is important to emphasize that PFS is an essential requirement for SDN. The lack of it can lead to information disclosure, i.e., reveal different aspects of the network's state and the controller's strategy (e.g., proactive or reactive flow setup).

Established KDC technologies like Kerberos have robust implementations
and are intensely used by industry, which makes
us consider the logical single-point-of-failure  they
present as moderate, and an acceptable option for the current state of
the art. However, and though, as we said, the KDC is out of the scope
of the paper, we present mitigation measures to achieve PFS in case of
compromise of the KDC. We also plan, as future work, to investigate
towards the development of SDN KDCs resilient to accidental and
malicious faults, drawing from fault and intrusion tolerance
techniques~\cite{Verissimo09his}.

On the devices side, we make no claim about their sheer resilience,
since this is largely dependent on vendors.  More precisely, when a
controller and/or a forwarding device is compromised, we consider that
the attacker is able to obtain all knowledge of the victim device(s),
including all stored secrets and the session status.  
However, it is our goal to guarantee the confidentiality of all past communications through measures that allow us to achieve perfect forward secrecy.



\section{iDVV: Keep It Simple and Secure}
\label{sec:simplicity}

Integrated device verification values (iDVVs) are sequentially
generated to protect and authenticate requests between two networking
devices.  The generator is conceived so that its output sequence has
the indistinguishability-from-random and determinism properties. In
consequence, the same sequence of random-looking secret values is
generated on both ends of the channel, allowing the safe decentralized
generation/verification of per-message keys at both ends.  However, if
the seed and key initial values and the state of the generator are kept
secret, there is no way an adversary can know, predict or generate an
iDVV.

In other words, an iDVV is a
unique secret value generated by a device A (e.g. a forwarding device),
which can be locally verified by another device B (e.g. a controller).
The iDVV generation is made flexible to serve the needs of SDN.  iDVVs
can therefore be generated: (a) on a per message basis; (b) for a
sequence of messages; (c) for a specific interval of time; and (d) for
one communication session.  The main advantages of iDVVs are their low
cost and the fact that they can be generated locally, i.e., without
having to establish any previous agreement.

Different from standard KDF algorithms such as HKDF, which assumes that 
keying material is not uniformly random or pseudorandom, our keying material 
(i.e. seed and key) are random symmetric secrets (each of size 256 bits), 
generated by the KDC, with high entropy. In such cases, a strong hash function 
can be safely used to derive a key (RFC 4880). As shown by the results in 
Section~\ref{sec:eval}, the iDVV generation is simpler and faster than standard 
KDF algorithm such as HKDF (RFC 5869) and similar solutions.

\subsection{iDVV bootstrap}
\label{sec:bootstrap}

As discussed before, the association between two SDN devices, e.g.,
forwarding device $f_i$ and controller $c_j$, happens through the help
of KDC, under the protection of the long-term secret keys obtained
from registration ($K_{kf}$, resp. $K_{kc}$).  The outcome of the
association protocol is the distribution of two random secrets to both
devices: a seed $seed_{ij}$, and an association key $key_{ij}$.  The
iDVV mechanism is bootstrapped by installing these two secret values
in both the controller and the switch, to animate the iDVV generation
algorithms, which we describe next.

Note that the set-up and generation of the iDVV values are performed
in a deterministic way, so that they can be done locally at both ends.
However, as iDVVs will be used as keys by cryptographic primitives
such as MAC or encryption functions, they have to be indistinguishable
from random.  Hashing primitives are natural choices for our
algorithms, since they provide indistinguishable-from-random values if
one or more of the input values are known only by the sender and the
receiver.  This explains why it is crucial that seed and association
key are sent encrypted and therefore known only to the communicating
devices.  Moreover, in order to prevent information leakage, all
variables $seed$, $key$, and $idvv$ in the algorithms below should
have the same length, which we chose to be 256 bits in our
design. This length is commonly considered robust, and the evaluation
in Section~\ref{sec:eval-crnd} confirms that.  
From our experiments
reported ahead in Section~\ref{performance}, the hashing primitive to
be used is SHA512, which yields 512 bits, of which we will use the
most-significant $q$ bits if we need to reduce the output length to $q$
(as recommended by \cite{rfc4880}). For example, we use the
most-significant 256 bits of the SHA512 output as the key
for symmetric ciphers.

The initial iDVV value is deterministically created at both ends of
the association between two devices\footnote{For readability, we omit
  the device-identifying subscripts in the variables.}, by calling
function \texttt{idvv\_init}, which performs hashing on the concatenation 
of the initial $seed$ and $key$, as illustrated by 
algorithm~\ref{alg:iDVVinit}. 
After set-up, the generator is ready for
first use, as described in the following section.

\begin{algorithm}[th]
	\caption{iDVV set-up}
	\label{alg:iDVVinit}
	\begin{algorithmic}[1]
	\State \texttt{idvv\_init()}
        \State \fourxspace \texttt{\textbf{idvv} $\gets$ H(seed $||$ key)}
	\end{algorithmic}
\end{algorithm}

\subsection{iDVV generation}

After the bootstrap with the initial $idvv$ value, the
\texttt{idvv\_next} function is invoked on-demand (again,
synchronously at both ends of the channel) to autonomously generate
authentication or encryption keys that will be used for securing the
communications, as illustrated by algorithm~\ref{alg:iDVVgen2g}.

The $key$ remains the only constant shared secret between the
devices. The $seed$ evolves to a new indistinguishable-from-random
value each time \texttt{idvv\_next} is invoked to generate a new iDVV.
The new seed is the outcome of a hashing primitive $H$
  over the current $seed$ and current $idvv$ (line 2).
The \textit{new} $idvv$, output of function \texttt{idvv\_next}, is
the outcome of a hashing primitive $H$ over the concatenation of the
\textit{new} $seed$ and association key $key$. 


\begin{algorithm}[th]
	\caption{iDVV generation}
	\label{alg:iDVVgen2g}
	\begin{algorithmic}[1]
	\State \texttt{idvv\_next()}
        \State \fourxspace \texttt{seed $\gets$ H(seed $||$ idvv)}    
        \State \fourxspace \texttt{\textbf{idvv} $\gets$ H(seed $||$ key)}
	\end{algorithmic}
\end{algorithm}

\subsection{iDVV synchronization}

The iDVV mechanism is agnostic w.r.t. secure communication protocols,
and can be used in a number of ways, in a number of protocols, as a
key-per-message or key-per-session, etc.
The only key issue about iDVV generation, is to keep it synchronized
in both extremes of the channel. So, we discuss recommendations in
this regard.

As a generic baseline robustness technique, communication should be
authenticated (encrypt-then-MAC recommended), such that any messages
failing crypto (decryption or MAC verification), can be simply
discarded and that fact handled by whatever existing error recovery
mechanisms. This brings in robustness against de-synchronization, or
malicious attacks, as we show below.

iDVVs can get out of sync for a number of reasons, like speed
differences, omission errors, or even DOS attacks. 
When de-synchronization happens, a baseline technique
consists of advancing the iDVV of the ``slower'' end, to catch up.
This lets us introduce another baseline robustness technique: when
say, $idvv^k$ is advanced to $idvv^l$ ($k < l$) to re-synchronize, and
the operation is not successful (crypto fails), the old $idvv^k$ is
restored, and the message motivating the recovery, is discarded.
This restoration does not affect the PFS of communications because
 the $idvv^k$ (or newer) has not yet been used to secure the traffic 
 between the two communicating devices.

Suppose an attacker can forge a re-synchronization request to claim
that it is in a future state (i.e. with a more advanced iDVV), and
fool the recipient to advance its iDVV to catch up: then the attacker
is able to play DoS attacks by keeping on asking all devices to
synchronize to an advanced iDVV. This is foiled by the first
robustness technique, since the attacker cannot mimic valid crypto, so
the message is discarded, and the second robustness technique ensures
that the node gets back to the original iDVV state.

Now we discuss some styles of using iDVVs, and possible protocol
classes they serve:

\textit{Simple iDVV -} used as is, works for lock-step, or
producer-consumer communication, where the advance is, respectively,
either round based, alternatively dictated by each end, or dictated by
the producer.  

If the channel is unreliable, packet losses may occur,
and then the receiver (R) gets out of sync and is not able to verify
the next received message from sender S.  
If the network has a bounded
omission degree (maximum number of consecutive omissions), say $Od$, R
can perform a simple recovery process: its iDVV is successively
advanced up to $Od+1$ times, until it is able to verify the incoming
message. If the process fails, the message is discarded and the iDVV
goes back to the original value (as per the techniques discussed
above).

If packet losses can be unexpectedly high, or both ends send
competitively and/or in a non-synchronized way, this algorithm is not
suitable.

\textit{Indexed iDVV -} iDVVs are indexed by the generation
number. Also, they are operated in "one key per direction" mode, i.e.,
at each end, one iDVV is generated for each communication direction.
This way, they support competitive, non-synchronized
correspondents. This mode also supports unreliable, connectionless
protocols like UDP.

%

Each iDVV generated is indexed by a sequence number (the initial iDVV
being $idvv^0$) and the sequence number is included in the message
where the respective $idvv$ is used.  This way, each receiving end
(this works in either direction, as we have two pairs of iDVVs) can
know the exact $idvv$ number that should be used and, for example,
detect and recover from omissions, by generating $idvv$'s the
necessary number of times to resynchronize. Again, the process is
robust: if it fails, the message is discarded and the iDVV goes back
to the original value.

\textit{Session iDVV -} iDVVs now mark sessions, inside which sets of
messages are sent that use crypto related to the current session iDVV.
It is quite suitable for example, for connection-oriented protocols.

Each $idvv^j$ is valid for the entire session $j$. A session may be a
standard, long-duration session a la SSL, or artificially short,
rolling session, for higher security, e.g. in a timed (e.g. 1-minute)
way.  Anyway, at the end of the session and start of the next one, the
$idvv^j$ is updated to $idvv^{j+1}$. 

Messages pertaining to a session $j$, labelled $(j)$, may all use the
same $idvv^j$ key. However, better can be done: inside a session,
rolling per-message keys may be created, based on $idvv^j$, for
example, $k_N=H(idvv^j || N)$, used for message labelled $(j,N)$, the
N-th message in the j-th session. 
Whenever a message with label $(j,N)$ is received, if $j$ is the
current session, then the device calculates the key $H(idvv^j || N)$
and decrypts or verifies this message.  Again, if the process fails,
or $j$ does not match, the message is discarded and the iDVV goes back
to the original value.

\subsection{iDVV implementation and application}


iDVVs require minimal resources, which means that they can be
implemented on any device, from a simple and very limited smart card
to most existing devices.  In other words, they are a simple and
viable solution that can be embedded in any networking device.  Just
three values per association have to be securely stored --- the seed,
the association key and the iDVV itself --- in order to use iDVV
continuously.  Furthermore, only hash functions, simple to implement
and with a very small code base, are required to generate iDVVs.  Such
kind of resource is already available on all networking devices that
support traditional network protocols and basic security mechanisms.

We advocate (and demonstrate in Section~\ref{sec:eval}) 
that iDVVs are inexpensive and, as a result, can
be used on a per-message basis to secure communication.
%
It is worth emphasizing that, from a security perspective,
one fresh iDVV per message makes it much harder for attacks such as key
recovery~\cite{Handschuh2008Key}, advanced side channel
attacks~\cite{Benoit2010side}, among other general HMAC
attacks~\cite{Kim2006On}, to succeed.  
In fact, the one-time key approach 
was initially used for generating MACs.
Yet, it was let aside (i.e. replaced by keys with a longer lifetime) due 
to performance reasons. 
However, as the iDVV generation has a low cost (see Section~\ref{sec:eval-perf}),
we incur in a lower penalty.

Finally, iDVVs can have further practical applications. For instance, the 
TLS handshake can be used to bootstrap the iDVV. After that, iDVVs 
can be used as session keys, i.e., in security mechanisms such as 
encrypt-then-MAC.

\section{On the cost of security}
\label{performance}

In this section we provide a quantitative analysis of the impact of
cryptographic primitives on control plane communication.
Although the number of use cases is expanding, SDN has been mainly
targeting data centers.  As such, SDN controllers have to be capable
of dealing with the challenging workloads of these large-scale
infrastructures.  In these environments new flows\footnote{In spite of
  the fact that there are several definitions of flow in
  SDN~\cite{kreutz2014sdnsurvey}, we equate SDN flow with TCP flow for
  the sake of simplicity.} can arrive at a given forwarding device
every \SI{10}{\micro\second}, with a great majority of mice traffic
lasting less than 100ms~\cite{Benson2010NTC}.  This means that current
data centers need to handle peak loads of tens of millions of new
flows/s.  The control plane has to meet both the network latencies and
throughputs required to sustain these high rates.
Current controllers are capable of achieving a throughput of up to 20M flows/s using TCP~\cite{kreutz2014sdnsurvey}. 

So any effort to systematically secure control plane communications
has to meet these challenges.
In the following we try to put the problem in perspective, by
analysing the effect of including even the most basic security
primitives to ensure authenticity, confidentiality and integrity when
considering peak loads of this magnitude.

We start by analyzing the latency impact of TLS, relative to TCP, and
then we focus on hashes and MACs as they are the essential primitives
for authenticity and integrity of communication.
To measure the latency of control plane communication\footnote{Time
  required to send a \ofpacketin message and receive a \offlowmod
  message without taking into account any further processing time of
  the controller.} we used Linux's resource usage system call
(\texttt{getrusage()}) to get the user CPU execution time.  This
function is commonly used to measure the performance of cryptographic
primitives.
Then, we compare the
performance of 50+ hashing and MAC primitives, including different
implementations such as those provided by OpenSSL (version 1.0.0) and
PolarSSL (version 1.3.9), two of the most widely used SSL libraries.
We evaluate these primitives using a hardware platform that includes
two quad-core Intel Xeon E5620 2.4GHz, with 2x4x256KB L2 / 2x12MB L3
cache, 32GB SDIMM at 1066MHz, with hyper-threading enabled and
overclocking and dynamic CPU frequency scaling disabled.  These
machines ran Ubuntu Server 14.04 LTS and were connected via Gigabit
Ethernet.  

\subsection{The cost of secure channels}
\label{sec:perfAssess}

Our first experiments assess the compared average latency of TCP and TLS on control plane communication.
We analyse the latency of connection setup and of OpenFlow \ofpacketin/\offlowmod messages.
The OpenFlow \ofpacketin message is used by switches to send packets to the controller (e.g. when there is no rule matching the packet received in the switch). 
\offlowmod messages allow the controller to modify the state of an OpenFlow switch.
One of the two nodes of the evaluation platform emulates the controller, whereas the other assumes the role of the forwarding devices.
The emulation removes the overhead specific to the controller's implementation, for instance. 
In practice, there is a huge performance gap among different controllers, most of which due to the chosen technologies and implementation details.
Similarly, the performance of switching devices varies also a lot due to implementation details.
To eliminate the implementation-specific performance penalty, we wrote a multi-threaded controller and forwarding devices that just send and receive \ofpacketin and \offlowmod messages.
This also means  that the controller sends \offlowmod messages in parallel to the forwarding devices.

\begin{figure}[ht]
   \centering
   \includegraphics[width=0.95\columnwidth]{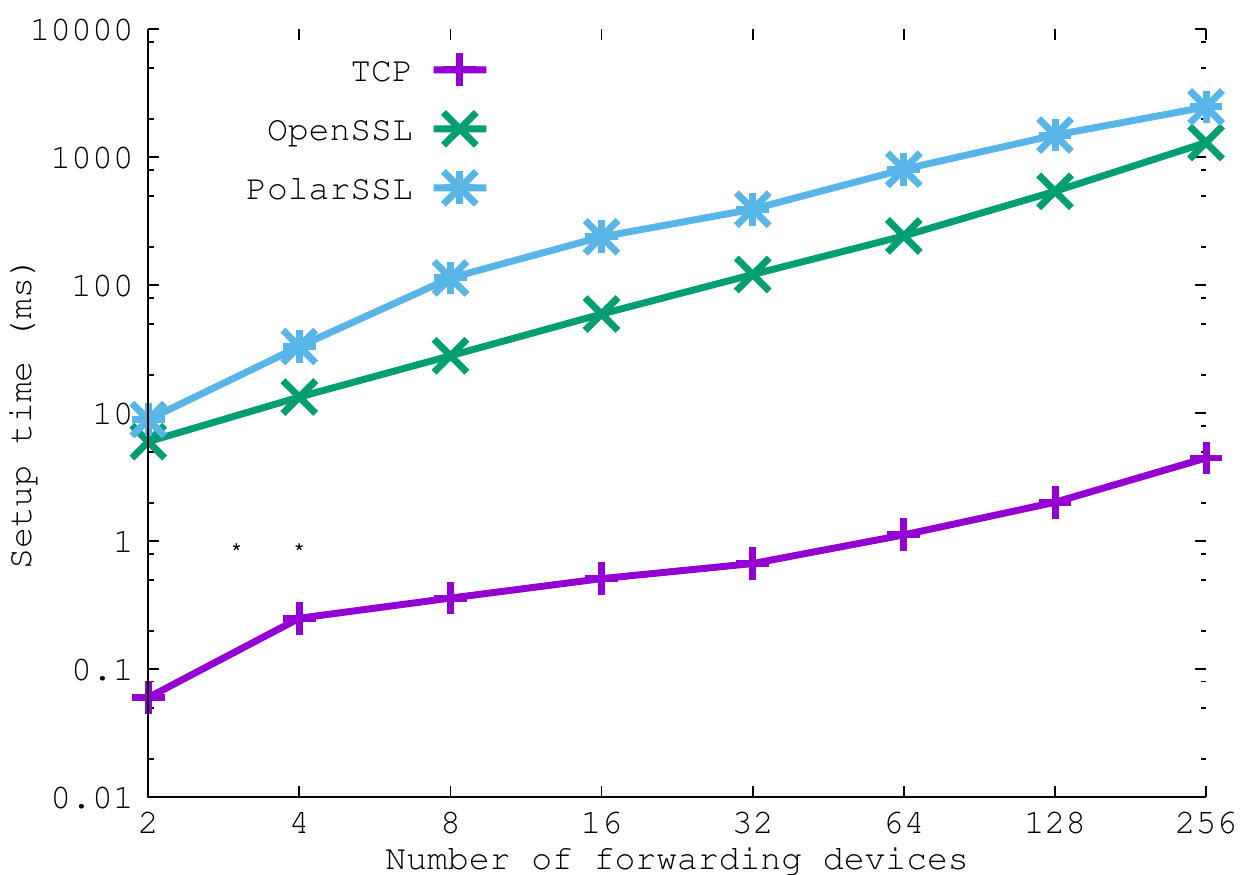}
   \caption{TCP and TLS connection setup times (in log scale)}
   \label{fig:perfConnectionSetup}
\end{figure}

\begin{figure}[ht]
   \centering
   \includegraphics[width=0.95\columnwidth]{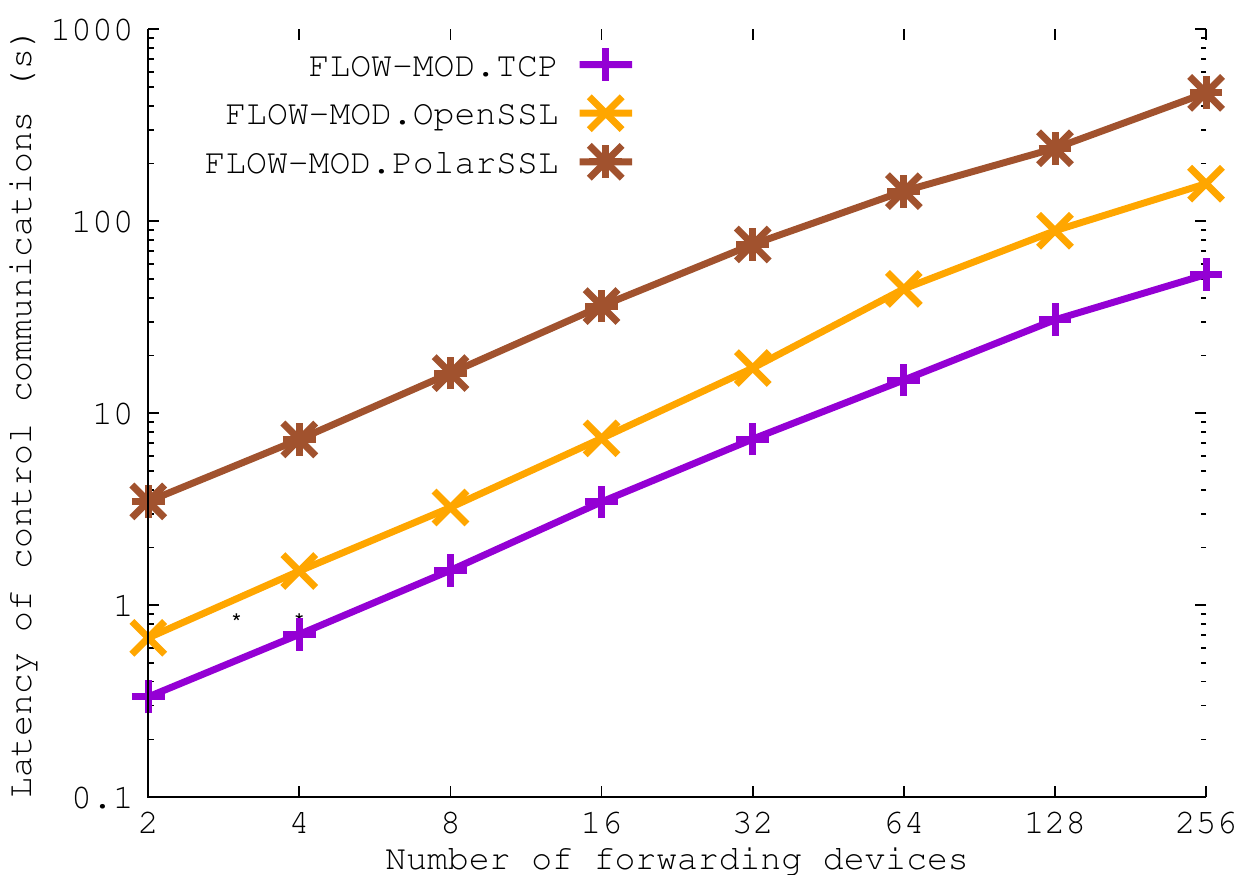}
   \caption{\offlowmod latency (in log scale)}
   \label{fig:perfOpenFlowInMod10000log}
\end{figure}

The emulated controllers and forwarding devices are implemented in C, using the OpenSSL and PolarSSL (a library used in systems from companies such as Gemalto, ARM, and Linksys) TLS implementations in their standard configuration (i.e. no library-specific optimizations were applied).
Figures~\ref{fig:perfConnectionSetup} and~\ref{fig:perfOpenFlowInMod10000log} show the median of the measured latency over 40k executions.
The standard deviation is below 3\% so we do not include it in the figures.

Figure~\ref{fig:perfConnectionSetup} shows the connection setup time (per forwarding device).
The higher costs of the two TLS implementations are due to the execution of a more elaborate handshake protocol between the devices.
While TCP uses a simple three-way handshake, TLS requires a nine message handshake for mutual authentication of the communicating entities.
As expected, the overhead increases with the number of forwarding devices.
Interestingly, our results also suggest that the choice of implementation has a non-negligible performance impact.
For connection setup, PolarSSL induces nearly twice the overhead of OpenSSL.

Although important, a high connection cost can be amortized by maintaining persistent connections. 
As such, the communications cost is usually considered more relevant.
Figure~\ref{fig:perfOpenFlowInMod10000log} shows the latency of 
\offlowmod messages (56 bytes, as specified in OpenFlow 1.4~\cite{ONF2013OpenFlow14}), averaged over 10k messages.
The results with \ofpacketin messages (32 bytes) were similar so we omit them for clarity. 
The costs of TCP, OpenSSL and PolarSSL grow nearly linearly with the number of forwarding devices.
OpenSSL latency is approximately 3x higher than TCP.
This is explained by the high overhead of cryptographic primitives, as we further analyse in the next section.
PolarSSL is significantly worse, increasing the latency by up to 7x when compared with TCP.

\textbf{Conclusions:} The main findings of this analysis can be
summarised in two points.  First, different implementations of TLS
present very different performance penalties.  Second, the additional
computation required by the cryptographic primitives used in TLS leads
anyway to a non-negligible performance penalty in the control plane.
In consequence, we turn to lightweight cryptographic libraries, such
as NaCl~\cite{Bernstein2012TSI} and
TweetNaCl~\cite{Bernstein2015TweetNaCl}, which are starting to be used
in different applications.  NaCl has been designed to be secure and to
be embedded in any system~\cite{Bacelar2013FV}, taking a clean slate
approach and avoiding most of the pitfalls of other libraries (e.g.
OpenSSL -- misuse issues).  First, it exposes a simple and high-level
API, with a reduced set of functions for each operation. Second, it
uses high-speed and highly-secure primitives, carefully implemented to
avoid side-channel attacks. 
Third, NaCl is less error-prone because low-security options are eliminated 
and it also provides a limited number of cryptographic primitives. In other 
words, users do not need deep knowledge regarding security to use it correctly. 
This is one of the major differences between it and other libraries such as OpenSSL. 
For instance, it has been recurrently shown that developers have been using 
OpenSSL in incorrect ways, leading to several security issues. Fourth, it has 
already been shown that secure and high-performance network protocols, 
outperforming OpenSSL, can be designed and implemented using 
NaCl~\cite{Petullo2013MMN}.

\subsection{A closer look at the cost of cryptography}
\label{sec:primitivesPerf}

To understand in more detail the cause of the previous
  findings we now perform a fine-grained analysis of two main classes
  of security primitives used in secure channel protocols: hashing and
  MAC.  

To measure the overhead of these primitives we disabled hyper-threading, in order to remove noise and randomness due to the implied resource sharing. 
As commodity switching devices do not implement direct cache access, we have ensured that the data to be hashed resides in main memory.
This avoids artificial performance boosts when operating on cached data\footnote{With cached data, we observed artificial gains of up to 20\% for hashing and of 12\% for MAC primitives.}. 
To mimic the behaviour of a switch, we circulated over an input buffer that is twice as large as the last-level cache (L3) to ensure that every read resulted in a cache miss.
The numbers in the following graphs represent the median of 1M executions, with a standard deviation below 3\%.

\begin{figure}[ht]
   \centering
   \includegraphics[width=0.95\columnwidth]{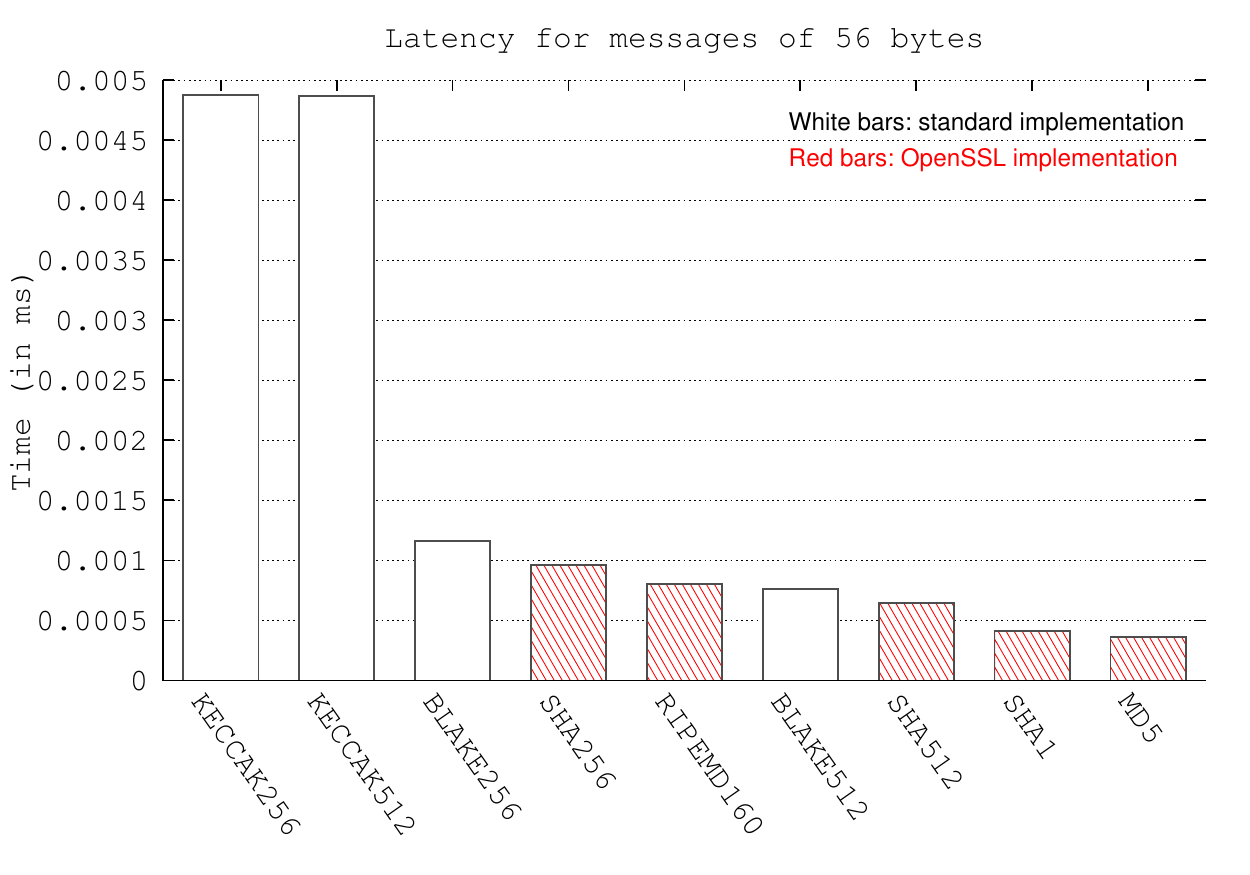}
   \caption{Hashing primitives}
   \label{fig:hash:xeonQplatform}
\end{figure}

We analyse the performance of nine hashing primitives.
The results are presented in Figure~\ref{fig:hash:xeonQplatform}.
The red bars represent primitives that are provided by OpenSSL, while white bars (BLAKE and KECCAK) indicate the original implementation of primitives that are not part of OpenSSL.
From Figure~\ref{fig:hash:xeonQplatform}, we observe that the primitives with smaller digest sizes (SHA-1 and MD5) achieve better performance, as expected. 
The stronger versions of the SHA and BLAKE families achieve comparable performance (slightly slower), with higher security guarantees.
Interestingly, SHA-512 outperforms SHA-256. 
This behavior is explained by the fact that on a 64-bit processor each round can process twice as much data (64-bit words instead of 32-bit words). However, SHA-256 is faster on a 32-bit processor.
In the case of KECCAK the difference in performance is due to the additional computational complexity of the mechanisms employed.
For instance, this solution requires 24 rounds of permutation on each compression step, while BLAKE requires up to 16 rounds.

\begin{figure}[ht]
   \centering
   \includegraphics[width=0.95\columnwidth]{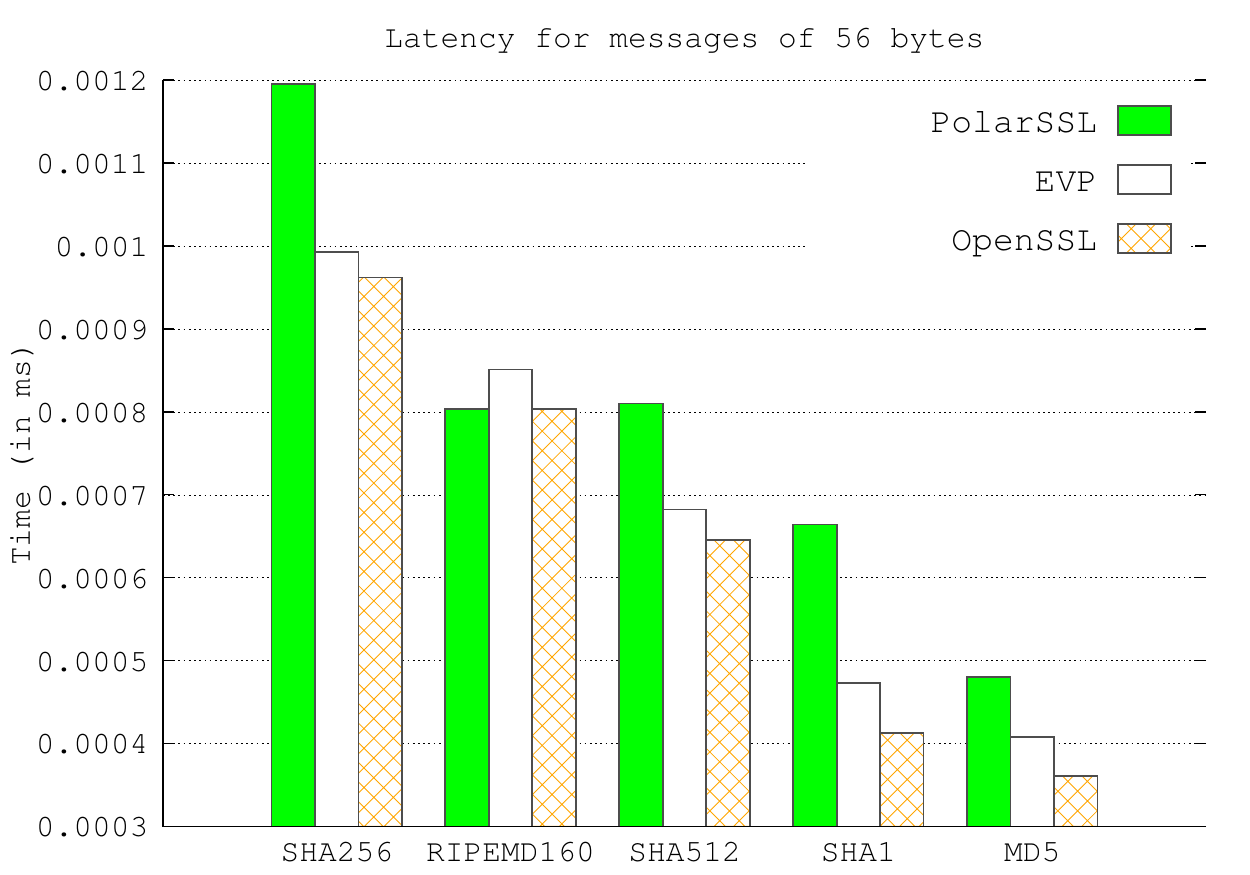}
   \caption{Implementations of hashing primitives}
   \label{fig:hash:xeonQplatformImpl}
\end{figure}

To understand the variance between different implementations, we present in Figure~\ref{fig:hash:xeonQplatformImpl} the costs of the five hashing primitives for which different implementations were available.
The OpenSSL implementation shows the best performance performance for hashing primitives.
With the exception of RIPEMD160, the PolarSSL implementation always presented higher message latencies. 
In addition to OpenSSL and PolarSSL, we included EVP, a library that provides a high-level interface to cryptographic functions.
Its main purpose is the ability to replace cryptographic algorithms without having to modify applications.
The added flexibility comes at a cost, as we can observe in the results.
The same OpenSSL primitives used through an EVP interface experience a penalty between 3\% and 15\%.

\begin{figure}[ht]
   \centering
   \includegraphics[width=0.95\columnwidth]{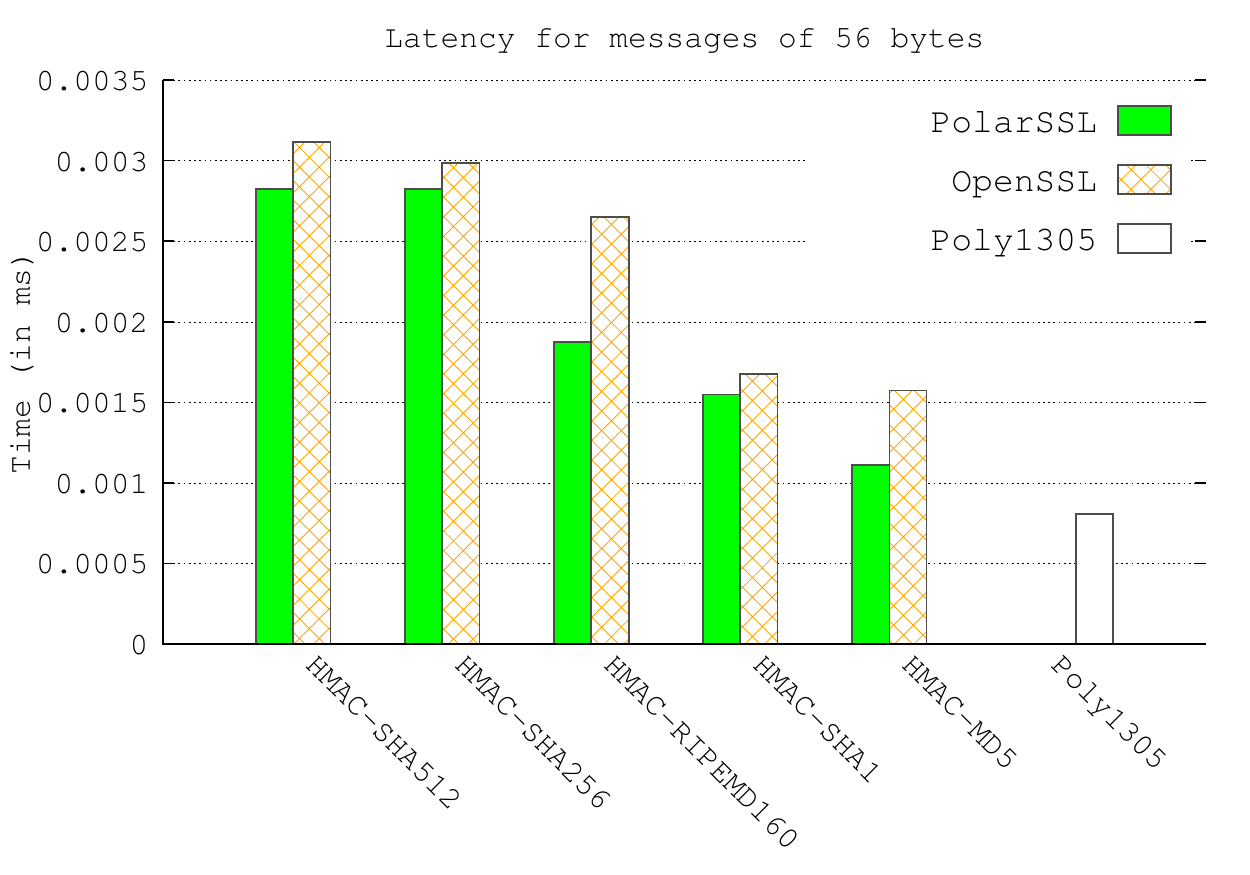}
   \caption{MAC primitives}
   \label{fig:hmac:xeonQplatform}
\end{figure}

Finally, Figure~\ref{fig:hmac:xeonQplatform} shows the results of the
latency analysis of six MAC primitives.  It is clear that Poly1305
outperformed all other primitives,
being approximately two times faster than OpenSSL's HMAC-SHA1, and
close to four times faster than HMAC-SHA512, for instance.  For MAC
primitives, the choice of specific implementations remains relevant.
Curiously, in this case the PolarSSL implementation always
outperformed the equivalent OpenSSL implementation.  The reason may
lie on the fact that OpenSSL does not provide native HMAC
implementations, but rather highly configurable HMACs through EVP
interfaces.  These primitives thus carry the overhead of EVP and the
extra costs of configurability.

\textbf{Conclusions:} From the results of
Figure~\ref{fig:hmac:xeonQplatform}, considering the MAC primitive with
best performance in the analysis (Poly1305 with 0.001ms per message),
around 20 dedicated cores are needed to compute a MAC in order to
maintain a rate of 20M flows/s.  To understand the importance of
judiciously selecting the security primitives implementation, the
HMAC-SHA512 OpenSSL (worst case performance in the analysis) would
require over three times more cores (up to 65) to compute MACs at
these rates.
From the hashing primitives analysis, we
conclude that SHA-512 performs best among the strong 
primitives (i.e. all except SHA1 and MD5), even better than SHA-256.
Concerning MAC primitives, the performance of HMAC-SHA512 disappoints,
and it is clear that Poly1305 outperformed all other primitives,
providing security with high speed and low per-message overhead.

In summary, our findings in this section indicate that (i) the
inclusion of cryptographic primitives results in a non-negligible
performance impact on the latency and throughput of the control plane;
and that (ii) a careful choice of the primitives used and their
respective implementations can significantly contribute to reduce this
performance penalty and enable feasible solutions in certain
scenarios.  Taking the outcome of our analysis into consideration, and 
given the benefits of NaCl described in Section~\ref{performance}, we
have selected the NaCl lightweight cryptographic library, and the MAC
and hash primitives with best performance -- Poly1305 and SHA512
OpenSSL -- as the baseline SC secure channel component technologies.
 NaCl is complemented in our architecture with the iDVV mechanism to generate crypto material (e.g. keys) used by NaCl ciphers.
Taken together they provide, as per our evaluation, the best trade-off
between security and performance for control plane communications in
SDN.  
We evaluate the overall result in the next section.

\section{iDVV evaluation}
\label{sec:eval}

\subsection{Performance}
\label{sec:eval-perf}

\begin{figure}[ht]
   \centering
   \includegraphics[width=0.95\columnwidth]{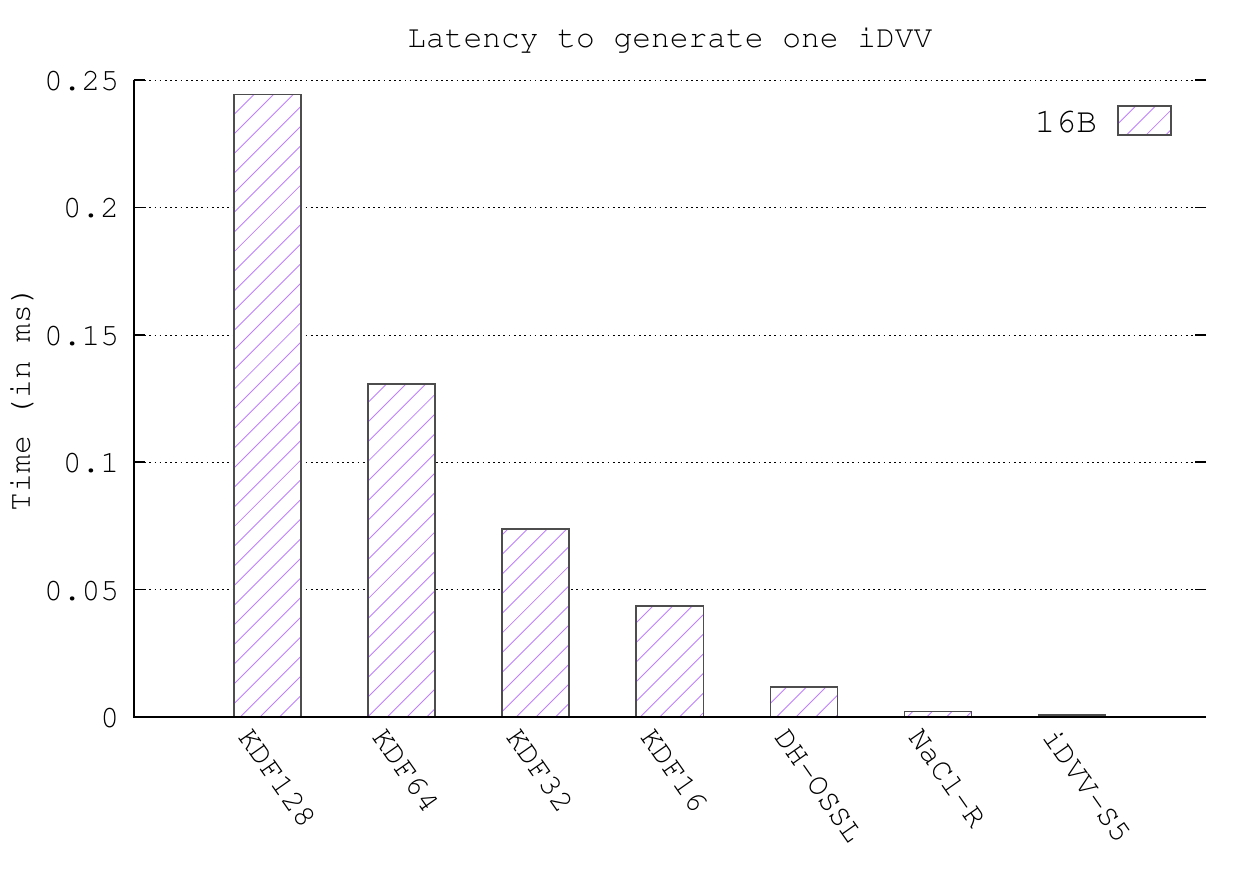}
   \caption{Latency to generate keys}
   \label{fig:results:iDVVsha512vsr}
\end{figure}

Figure~\ref{fig:results:iDVVsha512vsr} shows the performance of
different primitives for generating cryptographic material.  We
compare the iDVV generator using SHA512 (iDVV-S5), with an
implementation of a common key derivation function (KDFx) with
different values for the exponent $c$ (128, 64, 32, and 16,
respectively), the Diffie-Hellman implementation used by OpenSSL
(DH-OSSL), and the \texttt{randombytes()} function (NaCl-R) provided
by NaCl.  The latency of a KDF is very high, increasing linearly with
the number of iterations.  Our results for DH are compatible with
other publicly available performance measurements done on service
providers such as Amazon~\cite{Mavrogiannopoulos2011tpp}, showing a
latency several times higher than the iDVV generator.  The
\textit{randombytes()} primitive of NaCl, used to generate random
keys, is the second faster after iDVV, but still results 
a latency at least 2.6x higher.  NaCl-R's main latency lies on I/O operations required
to read the special random number generator device of the Linux
kernel, the \texttt{/dev/urandom}.  Last, but not least, it is worth
emphasizing that NaCl-R cannot be used for the same purposes of iDVVs,
since it only generates non-sequential random values, i.e., the values
would be different on both ends of the communication channel,
defeating our initial purpose.

\subsection{Security}
\label{sec:eval-corr}

Now, we provide a security analysis for the iDVV
Algorithms~\ref{alg:iDVVinit} and~\ref{alg:iDVVgen2g}, proving that
they provide indistinguishable-from-random and deterministic outputs.

\textbf{Algorithm~\ref{alg:iDVVinit} }

\begin{theo}
  \label{theo_init}
  If the initial values of seed and key are indistinguishable
  from random, then the resulting initial idvv (line 2) is
  indistinguishable from random.
\end{theo}

\begin{proofm}
  The $seed$ and $key$ are, by assumption (Section~\ref{sec:arch}) of
  the availability of robust sources of pseudo-random number
  generators in the central services (which generate the former),
  indistinguishable from random.  In consequence, and assuming that
  $H$ is a strong hash function, the output of H(seed $||$ key) will
  thus be indistinguishable from random.
\end{proofm}


\begin{theo}
  \label{theo2_init}
   Any execution of the function \texttt{ H(seed $||$ key)} with
   the same input values seed and key, produces the same output value
   (\texttt{idvv} in line 2).
\end{theo}

\begin{proofm}
Proof that Algorithm~\ref{alg:iDVVinit} is deterministic follows
trivially from the deterministic nature of hash functions.
\end{proofm}

\textbf{Algorithm~\ref{alg:iDVVgen2g} }

\begin{lem}
  \label{lem1_seed}
  If the seed and idvv are indistinguishable
  from random, then the resulting new seed (line 2) is
  indistinguishable from random.
\end{lem}

\begin{proofm}
  We start by proving the result of the run with the initial values of
  $seed$ and $idvv$. The initial $seed$ is, by assumption
  (Section~\ref{sec:arch}) of the availability of robust sources of
  pseudo-random number generators in the central services (which
  generate the former), indistinguishable from
  random. Theorem~\ref{theo_init} states that the initial $idvv$ is
  indistinguishable from random.
  In consequence, with a similar argumentation of the proof of
  Theorem~\ref{theo_init}, and assuming that $H$ is a strong hash
  function, the output of H(seed $||$ idvv) will be
  indistinguishable from random.

  Now we recurse the argumentation, to show that the proof is valid
  for any input values of $seed$ and $idvv$.
  A \textit{new} $seed$ was just proven to be indistinguishable from
  random.  A \textit{new} $idvv$ is proven below in
  Theorem~\ref{theo1_idvv} to be indistinguishable from
  random. Feeding these into the argumentation above, we generalise
  the proof $\forall$ $seed$ and $idvv$.
\end{proofm}


\begin{theo}
  \label{theo1_idvv}
  If the seed and key are indistinguishable from
  random, then the resulting new idvv (line 3) is indistinguishable from
  random.
\end{theo}

\begin{proofm}
Lemma~\ref{lem1_seed} establishes that $seed$, output by line 2 and
thus used as input in line 3, is indistinguishable from random.  The
$key$ is, by assumption of the availability of robust sources of
pseudo-random number generators in the central services (which
generate the former), indistinguishable from random.

We start by proving the result of the run with the initial value
$idvv$. Theorem~\ref{theo_init} states that the initial $idvv$ is
indistinguishable from random.
  In consequence, with a similar argumentation of the proof of
  Theorem~\ref{theo_init}, and assuming that $H$ is a strong hash
  function, the output of H(seed $||$ key) will be
  indistinguishable from random.

Now we recurse the argumentation, to show that the proof is valid for
any values of $idvv$.  Any \textit{new} $idvv$ was just proven to be
indistinguishable from random. In some next run, it will pair with
$key$, by nature indistinguishable from random, and with any new
$seed$, proven by Lemma~\ref{lem1_seed} to be indistinguishable from
random. Feeding these into the argumentation above, we generalise the
proof $\forall$ $key$, $seed$ and $idvv$.
  
In other words, the newly generated iDVV is an indistinguishable from
random value that can be safely used as an authentication or
authorization code, secret key, random nonce, and so forth.
\end{proofm}


\begin{lem}
  \label{lem2_idvv}
   Any execution of the function \texttt{ H(seed $||$ idvv)} with the
   same input values seed and idvv, produces the same output value
   (\texttt{seed} in line 2).
\end{lem}

\begin{proofm}
Proof that the function is deterministic follows
trivially from the deterministic nature of hash functions.
\end{proofm}

\begin{lem}
  \label{lem3_idvv}
   Any execution of the function \texttt{ H(seed $||$ key)} with the
   same input values seed and key produces the same output value
   (\texttt{idvv} in line 3).
\end{lem}

\begin{proofm}
Proof that the function is deterministic follows
trivially from the deterministic nature of hash functions.
\end{proofm}

\begin{theo}
  \label{theo2_idvv}
   Any execution of Algorithm~\ref{alg:iDVVgen2g} with the same input
   values seed, idvv and key produces the same output value
   (\texttt{idvv} in line 3).
\end{theo}

\begin{proofm}
Proof that Algorithm~\ref{alg:iDVVgen2g} is deterministic follows
trivially from Lemma~\ref{lem2_idvv} and~\ref{lem3_idvv}: since the
two functions are executed in a row, and the $seed$ output of line 2 
used as input in line 3 is
deterministic (Lemma~\ref{lem2_idvv}), it satisfies the conditions of
Lemma~\ref{lem3_idvv} for determinism.
\end{proofm}

\subsection{Perfect forward secrecy}
\label{sec-pfs}

In this section, we provide a discussion about the perfect forward
secrecy properties of our protocols, in face of compromise of any of
KDC, controller, forwarding device.  We re-state our goal in that
case: safeguard secrecy of past communications from the time the key
became active, to the time it became known to the attacker.

Note that when the assumed key distribution authority (e.g. the
Kerberos KDC) is compromised, then the attacker is able to obtain all
the shared secrets $K_{kc}$ (resp. $K_{kf}$) between the authority and
every controller (resp. every forwarding device). In this case, the
attacker would be able to decrypt the past communication that
delivered the initial $seed$ and $key$ to the associated devices, and
in consequence, decrypt past conversations, since the generation of
iDVVs is deterministic from the initial state (see \texttt{idvv\_init}
in Section~\ref{sec:bootstrap}).

Although providing secure and robust key distribution services is an
open challenge and orthogonal to this paper, we provide a simple
mechanism for providing PFS even when the authority is compromised. We
achieve it by updating the shared key (between the authority and
registered devices) each time a forwarding device is associated with a
controller. The key is updated as follows: $K_{kc}\gets H(K_{kc})$ and
$K_{kf}\gets H(K_{kf})$. This way, a shared key captured cannot
decrypt any past messages, since they have been encrypted with
previous generations of that key, which have been ``forgotten'' in the
system, given the irreversible nature of hashes.

As far as devices are concerned, when they are compromised, the
current values of $seed$, $key$ and $idvv$ are captured. Mote
that $seed$ is rolled forward every time a new iDVV is generated.  Only
$key$ stays as the original secret, but short of having as well the initial
$seed$ as sent at the end of the association procedure, the attacker
will also not be able to synthesize any past iDVVs since day one and
so, cannot also decrypt past conversations, achieving PFS, as we
sought.

As far as devices are concerned, when they are compromised, the 
current values of $seed$, $key$ and $idvv$ are captured. Note that 
$key$ stays as the original secret, but $seed$ is rolled forward every 
time a new iDVV is generated. So, the attacker will be unable to 
synthesize any past iDVVs since day one and so, cannot decrypt 
past conversations, achieving PFS, as we desired.


\subsection{Randomness}
\label{sec:eval-crnd}

We empirically assessed the quality and confidence of the iDVV
generator using two techniques.  First, we generated more than 200 billion
iDVVs to verify if there was any repetition, i.e., the same iDVV
generated more than once.  There was no a single repeated iDVV.  
This indicates that our solution is (indeed) suitable for short term iDVVs
(e.g. one per message).

Second, pseudorandom generators should be always empirically tested~\cite{Bassham2010SRS}.
Again, we used NIST's test suite~\cite{nist2017sts} to statistically
assess the confidence of the iDVV generator.  For the sake of our
tests, we generated 1M iDVVs of 64 bytes.  The file, containing 1M
iDVVs, was used as input for the test suite.  The streams of bits
corresponding to the iDVVs passed all tests, i.e., there was no single
failure.  This gives us a good level of confidence on the robustness
of the iDVV generator.

We also used \texttt{ent}~\cite{Walker2008ent}, which is a
pseudorandom number sequence test program, to evaluate the serial
correlation coefficient of our implementation.  While non-random and
predictable sequences of bytes have a serial correlation coefficient
of approximately 0.5 and 1.0, respectively, a random byte stream
should have a coefficient near to zero. Our implementation, featuring
SHA512, had a serial correlation coefficient of 0.0004.
Alternative implementations, using MD5 and SHA1, presented the worst case coefficients, as high as 0.035.
Typical pseudo-random functions or methods provided by a programming
language, such as \textit{rand()} from C and \textit{SecureRandom}
from Java, have a serial correlation of approximately 0.0148 and
0.0127, respectively. This shows us that SHA512 is indeed a strong
candidate to securely generate iDVVs.


\section{Discussion}

\subsection{On the cost of the infrastructure}
\label{sec:versus}

Our proposal compares well with traditional solutions such as EJBCA
(\url{http://www.ejbca.org/}) and OpenSSL, two popular implementations
of  PKI and TLS, respectively.

The first interesting take away is that our solution has nearly one
order of magnitude less LOC (85k) and uses four times less external
libraries and only four programming languages.  This makes a huge
difference from a security and dependability perspective.  For
instance, to formally prove more than 717k LOC (OpenSSL + EJBCA) is by
itself a tremendous challenge.  And it gets considerably worse if we
take into account eighty external libraries and eleven development
languages.  Moreover, it is worth emphasizing that libraries such as
OpenSSL suffer from different fundamental issues such as too many
legacy features accumulated over time, too many alternative modes as
result of tradeoffs made in the standardization, and too much focus
on the web and DNS names.

Second, OpenSSL is complex and highly configurable. This has been 
also the source of many security incidents, i.e., developers and users 
frequently use the library in an inappropriate way~\cite{Egele2013ESC,Fahl2012WEM}. 
It has also been shown that the
majority of the security incidents are still caused by errors and
misconfiguration of
systems~\cite{Verizon2015dbir,zhang2014mismanagement}.  
Lastly, recent
research has uncovered new vulnerabilities on TLS implementations~\cite{beurdouche2015messy}.

In contrast, our proposed architecture exhibits gains in both
performance and robustness, contributing to solving the dilemma we
enunciated in the introduction. By having less LOC, we significantly
reduce the threat surface -- by one order of magnitude -- and by
combining NaCl and the iDVV mechanism, we provide a potentially
equivalent level of security, but quite increased
performance/robustness product, as keys can be rolled even on a per
message basis.

\subsection{Size and complexity matter}
\label{sec:sizecomp}


The more complex the system, the higher the probability of having
vulnerabilities and hence a broader attack surface.  Nowadays, this is
still one of the major problems faced by the technology industry.
Specialized security reports have recurrently highlighted the
complexity and size of systems as one of the most important security
challenges~\cite{cisco2014asr}.  The time for re-thinking the security
of communication channels may have come, and that is also the position we
take in this paper.

Renowned cryptographers and security experts have been claiming that
simplicity is one of the keys in securing computer
systems~\cite{Bernstein2012TSI,stanford2015sce,Preneel2015ssa}.
In fact, the trusted computing community has been advocating simple
interfaces and concerned with the size and complexity of
components for a long time~\cite{heiser2012itt,raj2011credo}.

These positions have in essence been echoed in our KISS work (starting
with the name metaphor, \textbf{k}eep \textbf{i}t \textbf{s}imple,
\textbf{s}tupid).  
We methodically selected high performance MAC and
hashing primitives for KISS -- Poly1305 and SHA512 OpenSSL -- and
actually showed the penalty to be paid by less attentive
choices.
We also turned to lightweight but comparatively secure cryptographic
libraries for secure channel support, like NaCl.
NaCl was complemented in our architecture with the iDVV mechanism, to generate
secrets to be used for example by NaCl ciphers, again in a fast, very
simple and decentralized way.


\subsection{On the cost of iDVV}
\label{sec:iDVV:cost}


Similarly to iCVVs, iDVVs are a low overhead solution that requires
minimal resources.  This solution is thus feasible to be integrated
into compute-constrained devices as commodity switches.  Our
preliminary evaluation has revealed that the iDVV mechanism is faster
than traditional solutions, namely, the
key-exchange algorithms embedded in the OpenSSL implementation.
Considering a setup with 128 switching devices, sending 
\ofpacketin messages to and receiving 
\offlowmod messages from the controller
our results shows our
proposed solution (iDVV + NaCl's ciphers) to be more than 30\% faster
than an OpenSSL-based implementation using AES256-SHA (the most common
high performance cipher suite, used by IT companies such as Google,
Facebook, Microsoft, and Amazon).  
Importantly, we were able to
outperform OpenSSL-based deployments while still providing the same
security properties: authenticity, integrity, and confidentiality.  In
addition, we achieved this result not only while offering the same
properties, but also with stronger security guarantees: the tests were
made by generating one iDVV \emph{per packet}, while the OpenSSL-based
implementation uses a single key (for symmetric ciphering) for the
entire communication session.

\section{Related work}
\label{sec:related}

 There are several feasible attacks against the SDN control
  plane~\cite{hayward2016asurvey,kreutz2013HotSDN}.  Most of them
  explore vulnerabilities such as the lack of authentication,
  authorization and other essential security properties.  However,  
  almost no attention has been paid to the
  security requirements of control plane associations and
  communication between devices.  For instance, only recently, the use
  of secrecy through obscurity has been proposed to protect SDN
  controllers from DoS attacks~\cite{abdullaziz2016light}.  In this
  case, the switch authentication ID is hidden in a specific field in
  the IP protocol.  It is assumed that the devices share a look-up
  table and unique IDs.  However, in spite of being capable of
  mitigating DoS attacks, this technique does not address the security
  issues of control plane communications.

\section{Concluding remarks}
\label{sec:conclusion}

In this paper, we set out to explore and confirm our intuition for the
possible reasons behind a slower than expected adoption of security
mechanisms in SDN, and based on those findings, we proposed KISS, a
modular secure SDN control plane communications architecture.

We started by investigating the impact of essential cryptographic
primitives and TLS implementations on the control plane performance.  
We showed that whilst even the most basic security primitives add a non-negligible
degradation of performance, a judicious choice of these primitives and
their specific implementations can mitigate the penalty significantly.
This is particularly important for the typical SDN scenario that
resorts to commodity hardware, sometimes with modest computing
capabilities.

The second problem we explored in this paper was the complexity of the
centralized support infrastructure for authentication and key
distribution.  We proposed iDVV, a simple and robust decentralized
mechanism for generating and verifying the secrets necessary for
secure communications between network devices.  
As future work, we are also investigating the reduction of 
single-point-of-failure syndromes: architectures for SDN KDCs 
resilient to accidental and malicious faults, drawing from fault and 
intrusion tolerance techniques.

Our results are encouraging in terms of an increase of performance ---
30\% improvement over OpenSSL --- and robustness --- an order of
magnitude reduction in the number of LOC, and implied cyclomatic
complexity. This also means that formal verification is more
tractable, which is one of our future goals for iDVV, for instance. 

We believe that this is one first step towards lightweight but
effective security for control plane communication, and potentially
for SDN in general.  We make a ``call to arms'' to foster developments
on securing SDN communications without impairing performance, a
fundamental pre-condition for widespread adoption by future SDN
deployments.









%


\section*{Acknowledgment}

We would like to thank the anonymous reviewers of
IEEE Security \& Privacy and Christian Esteve Rothenberg and 
Marcus V\"{o}lp for the insightful comments.
It is worth mentioning that a short version of this report has been accepted for 
publication in IEEE Security \& Privacy. 

This work is partially supported by the
Fonds National de la Recherche Luxembourg (FNR) through PEARL grant
FNR/P14/8149128, by European Commission 
funds through the H2020 programme, namely by funding of the SUPERCLOUD project, ref.
H2020-643964, and by Portuguese national funds through Funda\c{c}\~{a}o
para a Ci\^{e}ncia e a Tecnologia (FCT), namely by funding of LaSIGE
Research Unit, ref. UID/CEC/00408/2013.

\IEEEtriggeratref{82}

\bibliographystyle{IEEEtran}
\bibliography{IEEEabrv,refs_kiss15_sp}

\end{document}